\documentclass[pra,twocolumn,floatfix,nofootinbib,showpacs]{revtex4}
\usepackage{amsmath}
\usepackage{graphicx}

\newcommand{\tprod}{\bigotimes}

\makeatletter
\def\btt#1{\texttt{\@backslashchar#1}}
\DeclareRobustCommand\bblash{\btt{\@backslashchar}}
\makeatother

\begin{document}

\title{Self-Protected Quantum Algorithms Based on Quantum State Tomography}

\author{Lian-Ao Wu}
\affiliation{ Center for Quantum Information and Quantum Control \\
Chemical Physics Theory Group, \\
Department of Chemistry, University of Toronto, \\
80 St. George Street, \\
Toronto, Ontario, Canada M5S 3H6}
\author{Mark S. Byrd}
\affiliation{Department of Physics and Department of Computer Science, 
Southern Illinois University, \\
Carbondale, IL 62901}

\begin{abstract}
Only a few classes of quantum algorithms are known which provide a speed-up
over classical algorithms. However, these and any new quantum algorithms
provide important motivation for the development of quantum computers. In
this article new quantum algorithms are given which are based on quantum
state tomography. These include an algorithm for the calculation of several
quantum mechanical expectation values and an algorithm for the determination
of polynomial factors. These quantum algorithms are 
important in their own right. However, it is remarkable that these quantum
algorithms are immune to a large class of errors. We describe these
algorithms and provide conditions for immunity.
\end{abstract}


\maketitle


\section{Introduction} 

There are only a few known quantum computing (QC) 
algorithms which provide a speed-up over their classical 
counterparts. The reasons for this are not completely clear 
\cite{Shor:03,Shoralgs:04}. However, those algorithms and the 
associated techniques for solving problems efficiently are quite 
valuable \cite{Shor:97,Grover:96,Feynman:QC}. For example, there 
are algorithms which belong to the same class as Shor's factoring
algorithm \cite{Shor:97} which enable the 
identification of a hidden abelian subgroup by
using a quantum Fourier transform. There is another 
set of algorithms belonging to the same 
class as Grover's search algorithm \cite{Grover:96} 
which can be applied to a wide class 
of problems where searching a solution set is the optimal known 
problem-solving strategy.  Yet another class consists of 
algorithms for simulating quantum systems.  Simulation algorithms 
can provide an exponential speed-up over any known 
classical algorithm for a variety of quantum systems 
\cite{Feynman:QC,Lloyd:96,Meyer:97,Boghosian:97a,Zalka:98,Abrams:97,Terhal:00,Freedman:02,Lidar:98RC,Ortiz:01,Wu/etal:BCS} 
and are very promising for many 
applications in the physical sciences. These include atomic, 
molecular, solid state, and nuclear simulations and do not 
necessarily require a fully scalable quantum computing device 
\cite{Jane:03}.  

To achieve the speed-ups promised by quantum computers, a reliable quantum
information processing device is required.  However, noise and imperfections 
still stand in our way. While the active strategies to prevent errors, such as
quantum error correcting codes 
\cite{Shor:95,Steane,Calderbank:96,Gottesman:97b} may, in principle, 
be universal as claimed, passive prevention methods have hardware resource
advantages. For example, decoherence-free subspaces (DFS) and noiseless 
subsystems (NS) 
\cite{Zanardi:97c,Duan:98,Lidar:PRL98,Knill:99a} 
are based on 
the symmetry of the system-bath interaction, so do not require active
detection and correction of errors.  Another passive 
technique, holonomic quantum computation, is
robust against stochastic errors in the control process 
\cite{Zanardi:99e,Pachos:99}. 
When conditions are appropriate, passive strategies
can be applied during the design of quantum algorithms.  

No matter which error prevention strategy is adopted however, 
it is widely believed that entanglement contributes to the 
errors in the system and is also the resource which 
is required to achieve the efficiencies promised by 
quantum computers.  Here we take advantage of quantum entanglement 
in an obvious way in order to provide algorithms which solve 
some problems in polynomial time on a QC device. 
Our quantum algorithms, some of which are able to 
calculate quantities which are now 
clearly out of reach for classical computing
devices, use quantum state tomography (QST) 
\cite{Vogel:89}.  

Our QST-based method complements the
scattering circuit method \cite{Somma:02,Paz:03,Alves:03,DAriano:04}, the
quantum phase estimation algorithms \cite{Kitaev:95,Cleve:98,Abrams:99} it
subsumes, and the adiabatic method we previously introduced for pairing 
Hamiltonians \cite{Wu/etal:BCS}. As our current method does, these other 
methods scale polynomially in the size of the input, and this is an 
exponential speedup over the best known classical algorithms for 
the same task.  However, our algorithms 
are exceptional since they are immune to 
a large class of errors and therefore share some features with 
the passive error protection methods of DFS/NS.  {\it Unlike this 
previous work, our emphasis is on the error resistance of the 
algorithms.}  

Specifically, we first describe an algorithm for the determination of
various types of observables that one may want to extract from a system
which is being used as a quantum simulator. Second, we provide a method for
the determination of the factors of a large polynomial using a quantum
computing device.  Our objectives are 1) to show why these algorithms 
are robust against errors and 2) how these and 
other algorithms can take advantage of such inherent robustness.


\section{Algorithm for obtaining the expectation values of an observable}

Let us first consider the simulation of a quantum system of $N$ subsystems
(e.g., particles). It is well known that Hamiltonians $H$ of the form $%
H=\sum_{k=1}^{L}H_{k}$, where $L$ is a polynomial in $N$, and such that
efficient quantum circuits exist for each term $H_{k}$ (e.g., when all $%
H_{k} $ have a tensor product structure or are simply sums of local terms),
generate unitaries $U=\exp (iHt)$ which can be polynomially simulated \cite%
{Nielsen:book}. Even random unitary matrices can be simulated polynomially 
\cite{Lloyd:03}. Let us denote by $O(N^{k})$ the simulation cost of such
efficiently simulatable unitaries, where $k$ is a fixed integer. This then
yields an efficient algorithm for obtaining the quantum state $\left\vert
\psi \right\rangle =U\left\vert 000...0\right\rangle _{N}$.

The most general operator of non-identical $d$-level particles (qudits), up
to two-body interactions, can be written as 
\begin{equation}
\mathcal{O}=\sum_{ij\alpha \beta \gamma \delta }O_{\alpha \beta ,\gamma
\delta }(ij)|\alpha _{i}\beta _{j}\rangle \langle \gamma _{i}\delta _{j}|,
\label{eq:O}
\end{equation}%
where $i,j$ label the $N$ subsystems and $\alpha ,\beta ,\gamma ,\delta \in
\{0,1,...,d-1\}$ the states of the qudits, and $\{\left\vert \alpha
_{i}\right\rangle \}$ is a basis for the Hilbert space of one qudit. For
example, this could be a Hamiltonian or a unitary gate.

We are interested in the expectation value $\left\langle \mathcal{O}%
\right\rangle =\langle \psi |\mathcal{O}\left\vert \psi \right\rangle $ in a
given quantum state $\left\vert \psi \right\rangle =U\left\vert \psi
(0)\right\rangle $ where $\left\vert \psi (0)\right\rangle $ is an initial
state. The existing classical algorithms for $\left\langle \mathcal{O}%
\right\rangle $ require the simulation of the unitary matrix $U$ with $%
d^{N}\times d^{N}$ independent elements. Clearly, the classical simulation
cost grows exponentially with the input. An efficient and general quantum
method for obtaining $\left\langle \mathcal{O}\right\rangle $ when $\mathcal{%
O}$ is unitary is the \textquotedblleft scattering
circuit\textquotedblright\ \cite{Paz:03}:\ one prepares an ancillary qubit in
the state $(|0\rangle +|1\rangle )/\sqrt{2}$, interacts the main system with
it using a controlled-$U$ operation, then measures the Pauli operator $%
\sigma ^{+}$ on the ancilla; this yields $\left\langle \mathcal{O}%
\right\rangle $ for qubits \cite{Somma:02,Paz:03,Alves:03,DAriano:04} (we
are unaware of a generalization of this method to qudits, though believe
this is possible). The scattering circuit method includes quantum phase
estimation algorithms \cite{Kitaev:95,Cleve:98,Abrams:99} as special cases;
its computational cost is $O(N^{k})$. Here we introduce a different general
method, based directly on QST \cite{Vogel:89}. Our method has a
computational cost that is higher by a factor of $O(N^{2})$ than the
scattering circuit, but it does not require an ancilla and, more
importantly, exhibits a remarkable inherent fault tolerance to decoherence
errors.

To this end it is convenient to re-express $\left\langle \mathcal{O}%
\right\rangle $ in a form relevant to QST. The two-qudit reduced density
matrix ${\rho }^{ij}$ is given by ${\rho }^{ij}=\sum_{m}\left\langle m|\psi
\right\rangle \left\langle \psi |m\right\rangle $, with $m$ running over all
the $d^{N-2}$ orthonormal basis vectors, excluding qudits $i$ and $j$. ${%
\rho }^{ij}$ is $d^{2}\times d^{2}$ dimensional, with elements ${\rho }%
_{\gamma \delta ,\alpha \beta }^{ij}=\left\langle \gamma _{i}\delta
_{j}\right\vert {\hat{\rho}}^{ij}\left\vert \alpha _{i}\beta
_{j}\right\rangle =\sum_{m}\left\langle \gamma _{i}\delta _{j}m|\psi
\right\rangle \left\langle \psi |m\alpha _{i}\beta _{j}\right\rangle
=\sum_{m}\left\langle \psi |m\alpha _{i}\beta _{j}\right\rangle \left\langle
\gamma _{i}\delta _{j}m|\psi \right\rangle =\left\langle \psi |\alpha
_{i}\beta _{j}\right\rangle \left\langle \gamma _{i}\delta _{j}|\psi
\right\rangle $, where we have used that $\left\langle \psi |m\alpha
_{i}\beta _{j}\right\rangle $ are $c$-numbers and $\sum_{m}\left\vert
m\right\rangle \left\langle m\right\vert =1$. Using Eq.~(\ref{eq:O}) we thus
have 
\begin{equation}
\left\langle \mathcal{O}\right\rangle =\sum_{ij\alpha \beta \gamma \delta
}O_{\alpha \beta ,\gamma \delta }(ij){\rho }_{\gamma \delta ,\alpha \beta
}^{ij}=\sum_{ij}\mathrm{Tr}(O(ij){\rho }^{ij}\mathbf{).}  \label{eq:<O>}
\end{equation}%
This expression implies an efficient quantum algorithm for $\left\langle 
\mathcal{O}\right\rangle $, as follows:\ (0) Classically calculate the $%
d^{2}\times d^{2}$ matrix elements $O_{\alpha \beta ,\gamma \delta }(ij)$
for all $N(N-1)/2$ distinct pairs of qudits, in the fixed basis $%
\{\left\vert \alpha _{i}\right\rangle \}$. (i) Propagate $\left\vert \psi
(0)\right\rangle $ to $\left\vert \psi \right\rangle $ using $U$, which can
be done in $O(N^{k})$ steps as noted above. (ii) Using QST find the $d^{4}-1$
real components of ${\rho }^{ij}$ (for a given pair of qudits $i,j$). (iii) Repeat steps
(i) and (ii) for all $N(N-1)/2$ distinct pairs of qudits. (iv) Repeat step
(iii) $M$ times, to obtain an estimate of $\left\langle \mathcal{O}%
\right\rangle $ with a precision (standard deviation) that scales as $1/%
\sqrt{M}$(central limit theorem). (iv) Classically evaluate $\sum_{ij}%
\mathrm{Tr}(O(ij){\rho }^{ij}\mathbf{)}$. The total simulation cost is $%
O(d^{4}MN^{k+2})$.  However, we may note that this might be improved 
using more recent QST methods \cite{Mohseni:06}. 

Note that this method can be generalized to the case of $n$-local
observables with many-body correlations. Specifically, any operator on $N$
qudits can be expressed as a linear combination of terms, each of which is a
tensor product of $N$ generalized Pauli matrices (e.g., the
\textquotedblleft very nice error operator basis\textquotedblright\ \cite%
{Ashikhmin:99a,Ashikhmin:99b}), 
where we include the $d\times d$ identity as a generalized
Pauli matrix. If each of these tensor products contains at most $n$
generalized Pauli matrices not equal to the identity then the operator is
said to be $n$-local. In the case of a $n$-local operator the obvious
generalization of Eq.~(\ref{eq:<O>}) is $\left\langle \mathcal{O}%
\right\rangle =\sum_{i_{1}i_{2}...i_{n}}\mathrm{Tr}(O(i_{1}i_{2}...i_{n}){%
\rho }^{i_{1}i_{2}...i_{n}}\mathbf{)}$, where ${\rho }^{i_{1}i_{2}...i_{n}}$
is the $d^{n}\times d^{n}$ dimensional reduced density matrix of particles $%
i_{1}i_{2}...i_{n}$. Its $d^{2n}-1$ real components can be obtained via QST,
again using a fixed number $M$ of copies of $\left\vert \psi
(0)\right\rangle $. This must be done for all $\left( 
\begin{array}{c}
N \\ 
n%
\end{array}%
\right) $ $n$-tuples of particles. Therefore the total computational cost of
our algorithm for the expectation value in the case of $n$-local observables
is $O(d^{2n}MN^{k+n})$. The measurement error $\epsilon =\langle (\mathcal{O}%
-\mathcal{O}_{\mathrm{est}})^{2}\rangle _{\mathrm{ave}}$ (where $\mathcal{O}%
_{\mathrm{est}}$ is the estimator employed and averaging is with respect to
the $M$ repetitions) satisfies the generalized uncertainty relation (derived
from the Cramer-Rao bound) \cite{Braunstein:94}: $\epsilon \Delta H\geq 1/(2%
\sqrt{M})$, where $\Delta H=(\langle H\rangle ^{2}-\langle H^{2}\rangle
)^{1/2}$ is the variance of $H$ on the input state $|\psi (0)\rangle $. This
bound is independent of $n$ but depends implicitly on $d$ through $\Delta H$%
. It is important to note that if $|\psi (0)\rangle $ is itself an entangled
state of $P$ identical copies then the measurement error can be reduced by a
factor of $P$ (the Heisenberg limit); the details and a general proof of
optimality of this bound, as well as its achievability, are discussed in 
\cite{Giovannetti:05}.

An important special case is when $\left\vert \psi \right\rangle $ is an
eigenstate of a Hamiltonian. The energy spectrum may then be found by
preparing a (complete) set of eigenstates $|\psi _{n}\rangle $ and measuring
the set of expectation values $\left\langle \mathcal{O}\right\rangle
_{n}=\left\langle \psi _{n}|H|\psi _{n}\right\rangle =E_{n}$. Let us comment
on precision issues in this context. Our QST-based method complements the
scattering circuit method \cite{Somma:02,Paz:03,Alves:03,DAriano:04} and the
quantum phase estimation algorithms \cite{Kitaev:95,Cleve:98,Abrams:99} it
subsumes, and the adiabatic method we previously introduced for pairing
Hamiltonians \cite{Wu/etal:BCS}. As in our current method, these other
methods scale polynomially in $N$, and this is commonly considered an
exponential speedup over the currently known best classical algorithms for
the same task. However, for error $\epsilon $ (defined above) the number of
digits of precision $l$ in the result is $l\symbol{126}\log (1/\epsilon )$,
and both the scattering circuit and the adiabatic methods require poly$%
(1/\epsilon )$ elementary steps to obtain this precision, due to the use of
the (quantum) Fourier transform at the measurement \cite{Brown:06}. In
contrast, an efficient algorithm would only require poly$(\log (1/\epsilon ))
$ number of steps. As observed by Brown et al. \cite{Brown:06}, while this
has no impact for fixed precision, the $1/\epsilon $ scaling does imply an
exponential scaling with the number of digits of precision. The origin of
the $l\symbol{126}\log (1/\epsilon )$ scaling is illucidated by Giovannetti
et al. \cite{Giovannetti:05}, who show that this scaling cannot be improved
even using entanglement. Namely, they show that entangled measurements do
not help, and the use of $P$ entangled input probes gives at most the
Heisenberg limit $\epsilon \symbol{126}1/P$, and on the other hand $l\symbol{%
126}\log P$. Thus $l\symbol{126}\log (1/\epsilon )$. While our QST\ based
method does not employ a Fourier transform at the measurement, the general
arguments used in \cite{Giovannetti:05} apply to QST\ as well, so that our
present algorithm does not improve on the precision issue. As discussed in 
\cite{Brown:06}, the origin of the poly$(1/\epsilon )$ number of steps is in
the use of the Trotter formula for the simulation of $U$. Use of the
Solovay-Kitaev theorem [which improves the Trotter poly$(1/\epsilon )$
scaling to $O(\log ^{2}(1/\epsilon ))$ scaling] does not help when a fault
tolerant implementation is considered, since the latter once again leads to
the poly$(1/\epsilon )$ scaling \cite{Brown:06}. However, it is important to
note that these general bounds do not preclude specific Hamiltonians from
being efficiently simulatable in terms of precision requirements; indeed the
exponential precision slow-down is avoided in Shor's algorithm due to the
manner in which modular exponentiation is carried out \cite{Shor:97}.
Another observation is that in some cases it is possible to prepare the
final state $|\psi \rangle $ by means other than quantum simulation, e.g.,
via cooling to the ground state, or via adiabatic evolution. There are
certainly examples where then reaching $|\psi \rangle $ from $|\psi
(0)\rangle $ requires poly$(\log (1/\epsilon ))$ steps.  


\section{Algorithm for obtaining the expectation values of an 
observable of a fermionic system} 

It follows from the Jordan-Wigner transformation \cite%
{Jordan/Wigner:26} that there is one-to-one correspondence between fermions
characterized by the fermionic creation and annihilation operations $%
c_{j}^{\dagger }$ and $c_{j}$, where $j$ denotes a fermionic mode, and
qubits: $c_{j}^{\dagger }\Leftrightarrow (-1)^{j-1}\left(
\tprod\nolimits_{l=1}^{j-1}\sigma _{l}^{z}\right) \sigma _{j}^{+}$ [$\sigma
^{\pm }=(\sigma ^{x}\pm i\sigma ^{y})/2$ and $\sigma ^{x,y,z}$ are the Pauli
matrices]. Therefore, the above algorithm can be applied to a fermionic
system. However, a one-body or two-body interactions of fermions usually
corresponds to a many-body interaction of qubits, for instance $%
c_{i}^{\dagger }c_{j}\Leftrightarrow (-1)^{i+j}\left(
\tprod\nolimits_{l=i}^{j-1}\sigma _{l}^{z}\right) \sigma _{i}^{z}...\sigma
_{j-1}^{z}\sigma _{i}^{+}\sigma _{j}^{-}$ where $j>i.$ Even so, it is clear
that obtaining the expectation value of an observable with many-body
correlation still requires only polynomial time. For example, measuring an
observable such as $\sigma _{i}^{x}\sigma _{j}^{x}\sigma _{i+1}^{z}...\sigma
_{j-1}^{z}$ can be accomplished efficiently, as long as the distance between
$i$ and $j$ is finite and independent of $N$.  In some cases 
\emph{partial} QST suffices to obtain a
desired expectation value. For example, consider the one-body Fermi operator 
$h=\sum \epsilon _{i}n_{i}$ ($n_{i}=c_{i}^{\dagger }c_{i}$), where we assume
the $\epsilon _{i}$ are known. Then 
\begin{eqnarray}
\!\!\!\phantom{\rangle}_{f}\!\left\langle \psi \right\vert h\left\vert \psi \right\rangle_{f}\!\!
&=&\!\! \sum \epsilon _{i}\,_{f}\left\langle \psi \right\vert n_{i}\left\vert
\psi \right\rangle _{f} \nonumber  \\
&\Leftrightarrow &\sum \epsilon _{i}\left\langle \psi \right\vert \frac{%
1-\sigma _{i}^{z}}{2}\left\vert \psi \right\rangle =\sum \epsilon
_{i}|c_{\alpha _{1}...0_{i}...\alpha _{N}}|^{2} \nonumber  \\
  &=&  \sum \epsilon _{i}\rho_{11}^{i},
\end{eqnarray}%
where $\left\vert \psi \right\rangle =\sum_{i=1}^{N}\sum_{\alpha
_{i}=0}^{1}c_{\alpha }|\alpha \rangle $ ($\alpha =\{\alpha _{1},...,\alpha
_{N}\}$) is an arbitrary pure state of $N$ qubits and $\rho _{11}^{i}=(%
\mathrm{Tr}_{j\neq i}\left\vert \psi \right\rangle \left\langle \psi
\right\vert )_{11}$, $j=1...N$. $\Leftrightarrow $ means that there
is a one-to-one correspondence between $\left\vert \psi \right\rangle_{f}$
expressed by Fermi creation operators on the vacuum state and 
$\left\vert \psi \right\rangle $ 
expressed by the superposition of computational bases 
\cite{Wu/Lidar:02b}.


\section{Inherent robustness}

The accuracy of a
usual quantum algorithm requires that the final wave function $\left| \psi
_{0}\right\rangle $ or density matrix be $\rho _{I}=\left| \psi
_{0}\right\rangle \left\langle \psi _{0}\right|$. In reality, due to errors,
the actual density matrix will be given by $\rho
_{A}=\sum_{k=0}^{2^{N}}p_{k}\left| \psi _{k}\right\rangle \left\langle \psi
_{k}\right| $ which is different from the ideal one $\rho _{I}.$ However as
long as the following relations are satisfied; 
\begin{equation}
\rho^{i} = \text{Tr}_{i}\rho_{I} = \text{Tr}_{i}\rho_{A} \;\;\;\text{or,}%
\;\;\; \rho^{ij} = \text{Tr}_{ij}\rho_{I} = \text{Tr}_{ij}\rho_{A}
\end{equation}
our algorithms will give the same results, where the subscripts $i$ and $j$
means trace over all degrees of freedom excluding $i$ and $j$. There are
only $3N$ or $15N^{2}$ constraints respectively. This implies that our
algorithms are much more fault-tolerant than a generic one. The reason that
the algorithms are more robust is that there are $2^{N}$ independent
coefficients in $\rho_{A}$. However, we only require that the above relation
holds independent of the other various parameters in the system.

As a motivational example, suppose an expected final state is $\left| \psi
_{0}\right\rangle =a\left| 00\right\rangle +b\left| 11\right\rangle ,$ but
due to dephasing errors, we actually get 
\begin{equation*}
\rho _{A}=|a|^{2}\left| 00\right\rangle \left\langle 00\right|
+|b|^{2}\left| 11\right\rangle \left\langle 11\right| +C\left|
00\right\rangle \left\langle 11\right| +C^{\ast }\left| 11\right\rangle
\left\langle 00\right|
\end{equation*}%
where $C$ is an arbitrary number and is zero when complete phase damping
occurs. No matter what value of $C$, 
\begin{equation*}
\rho ^{1}=\rho ^{2}=\left[ 
\begin{array}{cc}
|a|^{2} & 0 \\ 
0 & |b|^{2}%
\end{array}%
\right] .
\end{equation*}%
Therefore, dephasing does not affect the validity of our algorithm in this
case. The algorithms are, to some extent, self-protected.

To provide some general conditions under which our algorithms are robust,
let us start with some definitions. Let 
\begin{equation}
\rho = \rho_A\otimes \rho_B\otimes \rho_E,
\end{equation}
where $\rho_A$ is the subsystem we wish to study, $\rho_B$ is the rest of
our system, and $\rho_E$ is the density operator for the environment. We can
assume that each of these is a pure state and the whole system plus
environment is pure and initially completely separable. Now, let $U\otimes
I_E$ be the ideal unitary operation for our simulation algorithm and 
\begin{equation}
\rho_I = U\otimes I_E\rho U^\dagger \otimes I_E.
\end{equation}
Let $V$ be the non-ideal operation. We can write the condition for the
algorithm to give the same result for the expectation value of an operator $%
\mathcal{O}$ as $\mbox{Tr}(\mathcal{O}\rho^\prime) = \mbox{Tr}(\mathcal{O}%
W\rho^\prime W^\dagger), $ where $W=V(U^\dagger\otimes I_E)$. Let the basis
for the algebra of operators be traceless and Hermitian and represented by $%
\lambda^{(i)}_\alpha$, $i=1,2,3$, for subsystems $A,B,E$ respectively with $%
\alpha\in \{1,...,d^2-1\}$. Then we may write the density operator for the $%
ABE$ system as 
\begin{eqnarray}
\rho_{ABE}&=& I_{ABE}+\sum_ia_i\lambda^{(1)}_i +
I_A\otimes\sum_jb_j\lambda^{(2)}_j\otimes I_E  \notag \\
&& +I_{AB}\otimes \sum_kc_k\lambda^{(3)}_k.
\end{eqnarray}
If $\mathcal{O} = \vec{n}\cdot \vec{\lambda}^{(1)}$ (or $\mathcal{O} = \vec{n%
}\cdot \vec{\lambda}^{(1)}\otimes I_E$), then $\mathcal{O}$ acts as a
projector onto the subspace $A$ and the expectation value of $\mathcal{O}$
is 
\begin{equation}
\langle\mathcal{O}\rangle = \mbox{Tr}(\mathcal{O}W\rho^\prime W) = \mbox{Tr}(%
\vec{n}\cdot \vec{\lambda}^{(1)}W\rho^\prime W) .
\end{equation}
So if $W\rho^\prime W^\dagger$ has the form $\rho_{ABE}$, then $\langle%
\mathcal{O}\rangle = \vec{n}\cdot\vec{a},$ where $\vec{a} = \{a_1,a_2,
\dots\}$. Likewise, if 
\begin{equation}
\rho^\prime = I_{AB} +\sum_ia_i^I\lambda^{(1)}_i\otimes I_B +I_A \otimes
\sum_j b_j^I \lambda_j^{(2)},
\end{equation}
then $\mbox{Tr}(\mathcal{O}\rho^\prime) = \vec{n}\cdot\vec{a}^I. $
Therefore, for these to be equal, we require that $\vec{n}\cdot\vec{a} = 
\vec{n}\cdot\vec{a}^I. $ If we write $\vec{n}\cdot\vec{a}^I = |\vec{n}||\vec{%
a}^I|\cos \theta^I, $ and $\vec{n}\cdot\vec{a} = |\vec{n}||\vec{a}|\cos
\theta, $ then we need $|\vec{a}^I|\cos \theta^I = |\vec{a}|\cos \theta. $
For a two-state subsystem $A$, this leaves one degree of freedom, the little
group of the vector $\vec{a}$. This is stated in terms of the coherence
vector for a general expectation value for a $d$-state system.

We may also show that the robustness can be expressed in terms of the
expectation value of the operator $\mathcal{O}$ and completely positive (CP)
maps. Let us choose an initial density matrix $\rho$ which will be acted
upon by a CP map corresponding to the operator-sum decomposition with
operators $A_i$. We then want to find: 
\begin{equation}
\langle \mathcal{O} \rangle_1 = \mbox{Tr}(\mathcal{O}\sum_iA_i\rho
A_i^\dagger).
\end{equation}
Note that this can be written as 
\begin{equation}
\langle \mathcal{O} \rangle_1 = \mbox{Tr}(\sum_iA_i^\dagger \mathcal{O}A_i
\rho)
\end{equation}
so that the condition for the same result to be obtained from a different
set of operators $B_i$ is 
\begin{eqnarray}
0 &=& \langle \mathcal{O} \rangle_1- \langle \mathcal{O} \rangle_2  \notag \\
&=& \mbox{Tr}\left(\sum_i A_i^\dagger \mathcal{O}A_i \rho\right) - \mbox{Tr}%
\left(\sum_j B_j^\dagger \mathcal{O}B_j \rho\right)  \notag \\
&=& \mbox{Tr}\left[\left(\sum_i A_i^\dagger \mathcal{O}A_i - \sum_j
B_j^\dagger \mathcal{O}B_j \right) \rho\right].
\end{eqnarray}
Therefore, we may also say that the expectation value is invariant under
transformations which are comprised of the little group of $\mathcal{O}$.
This is true for both the unitary description above, as well as the
operator-sum decomposition.

Let us simplify to the case of a qubit. Letting $\mathcal{O}$ be traceless
and Hermitian and $B_j = \beta_j I + \vec{b}_j\cdot \vec{\sigma}$ and $A_i =
\alpha_i I + \vec{a}_i\cdot \vec{\sigma}$ we may obtain the relation 
\begin{eqnarray}
\sum_{i}A^\dagger_i \mathcal{O}A_i &=& \sum_{i} (iI(\vec{a}_i\times \vec{a}%
^*_i)\cdot \vec{n}  \notag \\
&&+ (|\alpha_i|^2 - \vec{a}\cdot \vec{a}^*)n_t\sigma_t  \notag \\
&&+ [\alpha_i(\vec{a}_i^*\times \vec{n})_t - \alpha_i^*(\vec{a}_i\times\vec{n%
})_t]\sigma_t  \notag \\
&&+[(\vec{a}_i^*\cdot\vec{n})a_{it}+(\vec{a}_i\cdot\vec{n}%
)a_{it}^*]\sigma_t),
\end{eqnarray}
where the sum over $t$ is implied. Simplifying further by letting $\mathcal{O%
} = \sigma_3$ and $\rho = (1/2)(I+\sigma_3)$, we can write the condition as 
\begin{eqnarray}
&&\sum_k[i(\vec{a}_k\times\vec{a}_k^*)_3 +
(|\alpha_k|^2-|a_{k1}|^2-|a_{k2}|^2 + |a_{k3}|^2)]  \notag \\
&& - \sum_j [i(\vec{b}_j\times\vec{b}_j^*)_3 -
(|\beta_j|^2-|b_{j1}|^2-|b_{j2}|^2 + |b_{j3}|^2)] = 0.  \notag
\end{eqnarray}
Note that $\sum_iA_i^\dagger A_i = I,$ implies $\sum_i(|\alpha_i|^2+\vec{a}%
_i\cdot\vec{a}_i^*) =1, $ and $\sum_i[\alpha_ia_{it}^* + \alpha_i^*a_{it}+i(%
\vec{a}_i\times\vec{a}_i^*)_t] = 0. $ So the result can be expressed in
terms of two equations 
\begin{eqnarray}
&&\sum_{k}[i(\vec{a}_{k}\times \vec{a}_{k}^{\ast
})_{3}-2(|a_{k1}|^{2}+|a_{k2}|^{2})]  \notag \\
&&-\sum_{j}[i(\vec{b}_{j}\times \vec{b}_{j}^{\ast
})_{3}+2(|b_{j1}|^{2}+|b_{j2}|^{2})] = 0.  \notag
\end{eqnarray}%
and 
\begin{eqnarray}
&&\sum_{k}[|\alpha _{k}|^{2}+|a_{k3}|^{2}+\alpha _{k}a_{k3}^{\ast }+\alpha
_{k}^{\ast }a_{k3}]  \notag \\
&&-\sum_{j}[|\beta _{j}|^{2}+|b_{j3}|^{2}+\beta _{j}a_{j3}^{\ast }+\beta
_{j}^{\ast }b_{j3}] =0.  \notag
\end{eqnarray}
To summarize, \textit{our simulation algorithms, based on quantum state
tomography and the expectation value of an operator, are immune to errors
which act as the little group of transformations of the initial density
operator or the operator for which we seek the expectation value.}


\section{Algorithm for factoring a polynomial} 

We now present one more
algorithm which can be implemented via state tomography and which is robust
against the aforementioned class of errors. Consider variables $x_{i},y_{i}$
, where $i=1,2,...,N,$ and a class of homogeneous functions spanned by the
set of products of $x_{i},y_{i}$. For instance when $N=2,$ the set is $%
x_{1}x_{2},x_{1}y_{2},y_{1}x_{2}$ and $y_{1}y_{2}$. There is a one-to-one
correspondence between this set and the computational basis for two qubits.
The linear combination of the set defines a class of homogeneous functions.
For instance, consider the two functions $x_{1}x_{2}+y_{1}y_{2}$ and $%
x_{1}y_{2}+y_{1}y_{2}.$ The former cannot be factored, while the latter can
be factored into the form $(x_{1}+y_{1})\times y_{2}.$ In some
circumstances, it may be easy to tell whether or not this can be factored,
if we know the concrete form of the homogeneous function. However, if a
homogeneous function contains many terms, in general it will become
difficult. Consider such a function derived from a matrix $U$ acting on a
basis set such as $x_{1}x_{2}x_{3}...x_{N}$ 
\begin{equation}
f_{N}(x_{i},y_{i})=Ux_{1}x_{2}x_{3}...x_{N}  \label{eq1}
\end{equation}%
where $U$ is a $2^{N}\times 2^{N}$ matrix. To represent the function, a
classical computer needs to handle $2\times 2^{N}\times 2^{N}$ independent
numbers in $2^{N}\times 2^{N}$ complex matrix elements of $U$ in order to
simulate it. We may assume $U$ is unitary so that it preserves the norm of
the function. However, this still requires a classical simulation of $%
2^{N}\times 2^{N}$ independent numbers in the matrix $U$ . When $N=300,$
approximately $10^{180}$ independent numbers must be handled.

Given a unitary matrix $U$ which could be a random unitary matrix, or the
quantum state $\left| f\right\rangle =U\left| 000...0\right\rangle _{N}$,
which can be simulated polynomially, say $N^{k}$, \cite{Lloyd:03} where $k$
is a fixed number, we will determine the presence of a factor $ax_{i}+by_{i}$%
. We first obtain the reduced density matrix of the ith qubit $\rho
^{i}=(1/2)(I+\vec{n}^{i}\cdot \vec{\sigma}^{i})$ by using quantum state
tomography, which requires need a fixed number, $M$, copies of $\left|
f\right\rangle$, as discussed above. Then, we calculate the von Neumann
entropy of $\rho ^{i}.$ If the entropy is zero, the ith qubit is separable
from the others, meaning that there is a factor $ax_{i}+by_{i}$ in $%
f_{N}(x_{i},y_{i})$ where $a=\sqrt{1+n_{z}^{i}}$ and $%
b=(n_{x}^{i}+in_{y}^{i})/\sqrt{1+n_{z}^{i}}$ given by the matrix elements of 
$\rho ^{i}.$ Otherwise, there is no such factor. The total number of steps
in the quantum procedure is $MN^{k}.$ The same procedure can be used to find
higher order factors. For example, a factor $%
ax_{i}x_{j}+bx_{i}y_{j}+cy_{i}x_{j}+dy_{i}y_{j}$ can be found by measuring
the reduced density matrix $\rho ^{ij}$ for the ith and jth qubits.

Furthermore, the generalization to many-qudit systems, (each subsystem has
an arbitrary dimension), and thus multivariate polynomials can be
accomplished by using the generalized coherence vector, or generalized Bloch
vector \cite{Mahler:book,Byrd/Khaneja:03,Kimura}. Let $\lambda^r_i \otimes
\mu^s_j$ be a Hermitian basis for a system of coupled qudits, with arbitrary
dimensions for all components. Let $\lambda^r_i$ form a basis for the ith
subsystem with Tr$(\lambda^r_i\lambda^t_i) =2\delta_{rt}$ and $\mu^s_j$ a
basis for the rest of the system. Then, given the state $U|000...0\rangle$
for the whole system, the corresponding reduced density matrix for the ith
subsystem has $\rho=(1/d)({%
\mathchoice {\rm {1\mskip-4.5mu l}} {\rm
{1\mskip-4.5mu l}} {\rm {1\mskip-3.8mu l}} {\rm {1\mskip-4.3mu l}}}+\vec{m}%
\cdot\vec{\lambda})$. The conditions for the system to be factorisable with
respect to the ith subsystem is that $\vec{m}\cdot\vec{m} = N(N-1)/2$ and $%
d^{rst}_i m_r m_s N(N-1)/(2N-4) = m_t$, with $d^{rst}_i=(N/4)\mbox{Tr}%
(\{\lambda^r_i,\lambda^s_i\}\lambda_i^t)$. As usual, $\{\cdot,\cdot\}$
denotes the anti-commutator. These conditions indicate that the qudit is in a
pure state and thus has zero entropy. Using the Hermitian basis for the
operators in this protocol provides an explicit measurement basis for the
identification of the reduced density matrices.


\section{Conclusions}

In this paper, we introduced quantum algorithms
based on quantum state tomography. The simulation algorithms are clearly
polynomial while the best known classical counterparts of the 
simulation algorithms are exponential. We suspect that the 
polynomial factoring algorithm is also more
efficient although we have not proved this generally. Certainly in the case
that the unitary $U$ must be simulated, we achieve an exponential speed-up.
We emphasize that the algorithms are, to a large
degree, self-protected against a large class of errors. This work brings
together two important aspects in quantum information science, algorithms
and passive correction. We expect that the family of quantum 
algorithms which are
error-avoiding algorithms should receive much more attention in future
studies of quantum algorithms.



\begin{thebibliography}{47}
\expandafter\ifx\csname natexlab\endcsname\relax\def\natexlab#1{#1}\fi
\expandafter\ifx\csname bibnamefont\endcsname\relax
  \def\bibnamefont#1{#1}\fi
\expandafter\ifx\csname bibfnamefont\endcsname\relax
  \def\bibfnamefont#1{#1}\fi
\expandafter\ifx\csname citenamefont\endcsname\relax
  \def\citenamefont#1{#1}\fi
\expandafter\ifx\csname url\endcsname\relax
  \def\url#1{\texttt{#1}}\fi
\expandafter\ifx\csname urlprefix\endcsname\relax\def\urlprefix{URL }\fi
\providecommand{\bibinfo}[2]{#2}
\providecommand{\eprint}[2][]{\url{#2}}

\bibitem[{\citenamefont{{P.W. Shor}}(2003)}]{Shor:03}
\bibinfo{author}{\bibnamefont{{P.W. Shor}}}, \bibinfo{journal}{{J. ACM}}
  \textbf{\bibinfo{volume}{50}}, \bibinfo{pages}{{87}} (\bibinfo{year}{2003}).

\bibitem[{\citenamefont{{P. Shor}}(2004)}]{Shoralgs:04}
\bibinfo{author}{\bibnamefont{{P. Shor}}}, \bibinfo{journal}{Qu. Inf. Proc.}
  \textbf{\bibinfo{volume}{3}} (\bibinfo{year}{2004}).

\bibitem[{\citenamefont{{P.W. Shor}}(1997)}]{Shor:97}
\bibinfo{author}{\bibnamefont{{P.W. Shor}}}, \bibinfo{journal}{SIAM J. on
  Comp.} \textbf{\bibinfo{volume}{26}}, \bibinfo{pages}{1484}
  (\bibinfo{year}{1997}).

\bibitem[{\citenamefont{{L.K. Grover}}(1996)}]{Grover:96}
\bibinfo{author}{\bibnamefont{{L.K. Grover}}}, in
  \emph{\bibinfo{booktitle}{{Proceedings of the 28th Annual ACM Symposium on
  the Theory of Computing}}} (\bibinfo{publisher}{ACM}, \bibinfo{address}{{New
  York, NY}}, \bibinfo{year}{1996}), p. \bibinfo{pages}{212}.

\bibitem[{\citenamefont{{R.P. Feynman}}(1982)}]{Feynman:QC}
\bibinfo{author}{\bibnamefont{{R.P. Feynman}}}, \bibinfo{journal}{Intl. J.
  Theor. Phys.} \textbf{\bibinfo{volume}{21}}, \bibinfo{pages}{467}
  (\bibinfo{year}{1982}).

\bibitem[{\citenamefont{{S. Lloyd}}(1996)}]{Lloyd:96}
\bibinfo{author}{\bibnamefont{{S. Lloyd}}}, \bibinfo{journal}{Science}
  \textbf{\bibinfo{volume}{273}}, \bibinfo{pages}{1073} (\bibinfo{year}{1996}).

\bibitem[{\citenamefont{{D.A. Meyer}}(1997)}]{Meyer:97}
\bibinfo{author}{\bibnamefont{{D.A. Meyer}}}, \bibinfo{journal}{Phys. Rev. E}
  \textbf{\bibinfo{volume}{55}}, \bibinfo{pages}{{5261}}
  (\bibinfo{year}{1997}).

\bibitem[{\citenamefont{{B.M. Boghosian and W. Taylor}}(1998)}]{Boghosian:97a}
\bibinfo{author}{\bibnamefont{{B.M. Boghosian and W. Taylor}}},
  \bibinfo{journal}{Phys. Rev. E} \textbf{\bibinfo{volume}{57}},
  \bibinfo{pages}{54} (\bibinfo{year}{1998}).

\bibitem[{\citenamefont{{C. Zalka}}(1998)}]{Zalka:98}
\bibinfo{author}{\bibnamefont{{C. Zalka}}}, \bibinfo{journal}{Proc. Roy. Soc.
  London Ser. A} \textbf{\bibinfo{volume}{454}}, \bibinfo{pages}{313}
  (\bibinfo{year}{1998}).

\bibitem[{\citenamefont{{D.S. Abrams and S. Lloyd}}(1997)}]{Abrams:97}
\bibinfo{author}{\bibnamefont{{D.S. Abrams and S. Lloyd}}},
  \bibinfo{journal}{Phys. Rev. Lett.} \textbf{\bibinfo{volume}{79}},
  \bibinfo{pages}{2586} (\bibinfo{year}{1997}).

\bibitem[{\citenamefont{{B.M. Terhal}}(2000)}]{Terhal:00}
\bibinfo{author}{\bibnamefont{{B.M. Terhal}}}, \bibinfo{journal}{Phys. Lett. A}
  \textbf{\bibinfo{volume}{{271}}}, \bibinfo{pages}{{319}}
  (\bibinfo{year}{2000}).

\bibitem[{\citenamefont{{M.H. Freedman,A. Kitaev and Z.
  Wang}}(2002)}]{Freedman:02}
\bibinfo{author}{\bibnamefont{{M.H. Freedman,A. Kitaev and Z. Wang}}},
  \bibinfo{journal}{Commun. Math. Phys.} \textbf{\bibinfo{volume}{227}},
  \bibinfo{pages}{587} (\bibinfo{year}{2002}).

\bibitem[{\citenamefont{{D.A. Lidar and H. Wang}}(1998)}]{Lidar:98RC}
\bibinfo{author}{\bibnamefont{{D.A. Lidar and H. Wang}}},
  \bibinfo{journal}{Phys. Rev. E} \textbf{\bibinfo{volume}{59}},
  \bibinfo{pages}{2429} (\bibinfo{year}{1998}), \bibinfo{note}{eprint
  quant-ph/9807009.}

\bibitem[{\citenamefont{{G. Ortiz, J. E. Gubernatis, E. Knill and R.
  Laflamme}}(2001)}]{Ortiz:01}
\bibinfo{author}{\bibnamefont{{G. Ortiz, J. E. Gubernatis, E. Knill and R.
  Laflamme}}}, \bibinfo{journal}{Phys. Rev. A} \textbf{\bibinfo{volume}{64}},
  \bibinfo{pages}{{022319}} (\bibinfo{year}{2001}).

\bibitem[{\citenamefont{{L.-A. Wu, M.S. Byrd and D.A.
  Lidar}}(2002)}]{Wu/etal:BCS}
\bibinfo{author}{\bibnamefont{{L.-A. Wu, M.S. Byrd and D.A. Lidar}}},
  \bibinfo{journal}{Phys. Rev. Lett.} \textbf{\bibinfo{volume}{89}},
  \bibinfo{pages}{{057904}} (\bibinfo{year}{2002}).

\bibitem[{\citenamefont{{E. Jane, G. Vidal, W. D\"ur, P. Zoller, J.I.
  Cirac}}(2003)}]{Jane:03}
\bibinfo{author}{\bibnamefont{{E. Jane, G. Vidal, W. D\"ur, P. Zoller, J.I.
  Cirac}}}, \bibinfo{journal}{Qu. Inf. \& Comp.} \textbf{\bibinfo{volume}{3}}
  (\bibinfo{year}{2003}).

\bibitem[{\citenamefont{{P.W. Shor}}(1995)}]{Shor:95}
\bibinfo{author}{\bibnamefont{{P.W. Shor}}}, \bibinfo{journal}{Phys. Rev. A}
  \textbf{\bibinfo{volume}{52}}, \bibinfo{pages}{2493} (\bibinfo{year}{1995}).

\bibitem[{\citenamefont{{A. Steane}}(1998)}]{Steane}
\bibinfo{author}{\bibnamefont{{A. Steane}}}, \bibinfo{journal}{{Rep. Prog.
  Phys.}} \textbf{\bibinfo{volume}{61}}, \bibinfo{pages}{117}
  (\bibinfo{year}{1998}).

\bibitem[{\citenamefont{{A.R. Calderbank and P.W. Shor}}(1996)}]{Calderbank:96}
\bibinfo{author}{\bibnamefont{{A.R. Calderbank and P.W. Shor}}},
  \bibinfo{journal}{Phys. Rev. A} \textbf{\bibinfo{volume}{54}},
  \bibinfo{pages}{1098} (\bibinfo{year}{1996}).

\bibitem[{\citenamefont{{D. Gottesman}}(1997)}]{Gottesman:97b}
\bibinfo{author}{\bibnamefont{{D. Gottesman}}}, Ph.D. thesis,
  \bibinfo{school}{California Institute of Technology},
  \bibinfo{address}{Pasadena, CA} (\bibinfo{year}{1997}), \bibinfo{note}{eprint
  quant-ph/9705052}.

\bibitem[{\citenamefont{{P. Zanardi and M. Rasetti}}(1997)}]{Zanardi:97c}
\bibinfo{author}{\bibnamefont{{P. Zanardi and M. Rasetti}}},
  \bibinfo{journal}{Phys. Rev. Lett.} \textbf{\bibinfo{volume}{79}},
  \bibinfo{pages}{3306} (\bibinfo{year}{1997}).

\bibitem[{\citenamefont{{L.-M Duan and G.-C. Guo}}(1998)}]{Duan:98}
\bibinfo{author}{\bibnamefont{{L.-M Duan and G.-C. Guo}}},
  \bibinfo{journal}{Phys. Rev. A} \textbf{\bibinfo{volume}{57}},
  \bibinfo{pages}{737} (\bibinfo{year}{1998}).

\bibitem[{\citenamefont{{D.A. Lidar, I.L. Chuang and K.B.
  Whaley}}(1998)}]{Lidar:PRL98}
\bibinfo{author}{\bibnamefont{{D.A. Lidar, I.L. Chuang and K.B. Whaley}}},
  \bibinfo{journal}{Phys. Rev. Lett.} \textbf{\bibinfo{volume}{81}},
  \bibinfo{pages}{2594} (\bibinfo{year}{1998}).

\bibitem[{\citenamefont{{E. Knill, R. Laflamme and L.
  Viola}}(2000)}]{Knill:99a}
\bibinfo{author}{\bibnamefont{{E. Knill, R. Laflamme and L. Viola}}},
  \bibinfo{journal}{Phys. Rev. Lett.} \textbf{\bibinfo{volume}{84}},
  \bibinfo{pages}{2525} (\bibinfo{year}{2000}).

\bibitem[{\citenamefont{{P. Zanardi and M. Rasetti}}(1999)}]{Zanardi:99e}
\bibinfo{author}{\bibnamefont{{P. Zanardi and M. Rasetti}}},
  \bibinfo{journal}{Phys. Lett. A} \textbf{\bibinfo{volume}{264}},
  \bibinfo{pages}{{94}} (\bibinfo{year}{1999}).

\bibitem[{\citenamefont{{J. Pachos, P. Zanardi and M.
  Rasetti}}(1999)}]{Pachos:99}
\bibinfo{author}{\bibnamefont{{J. Pachos, P. Zanardi and M. Rasetti}}},
  \bibinfo{journal}{Phys. Rev. A} \textbf{\bibinfo{volume}{61}},
  \bibinfo{pages}{{010305(R)}} (\bibinfo{year}{1999}).

\bibitem[{\citenamefont{{K. Vogel and H. Risken}}(1989)}]{Vogel:89}
\bibinfo{author}{\bibnamefont{{K. Vogel and H. Risken}}},
  \bibinfo{journal}{Phys. Rev. A} \textbf{\bibinfo{volume}{40}},
  \bibinfo{pages}{2847} (\bibinfo{year}{1989}).

\bibitem[{\citenamefont{{R. Somma, G. Ortiz, J. E. Gubernatis, E. Knill, and R.
  Laflamme}}(2002)}]{Somma:02}
\bibinfo{author}{\bibnamefont{{R. Somma, G. Ortiz, J. E. Gubernatis, E. Knill,
  and R. Laflamme}}}, \bibinfo{journal}{Phys. Rev. A}
  \textbf{\bibinfo{volume}{65}}, \bibinfo{pages}{{042323}}
  (\bibinfo{year}{2002}).

\bibitem[{\citenamefont{{J.P. Paz, A. Roncaglia}}(2003)}]{Paz:03}
\bibinfo{author}{\bibnamefont{{J.P. Paz, A. Roncaglia}}},
  \bibinfo{journal}{Phys. Rev. A} \textbf{\bibinfo{volume}{68}},
  \bibinfo{pages}{{052316}} (\bibinfo{year}{2003}).

\bibitem[{\citenamefont{{C.M. Alves, P. Horodecki, D.K.L. Oi, L. C. Kwek and
  A.K. Ekert}}(2003)}]{Alves:03}
\bibinfo{author}{\bibnamefont{{C.M. Alves, P. Horodecki, D.K.L. Oi, L. C. Kwek
  and A.K. Ekert}}}, \bibinfo{journal}{Phys. Rev. A}
  \textbf{\bibinfo{volume}{68}}, \bibinfo{pages}{{032306}}
  (\bibinfo{year}{2003}).

\bibitem[{\citenamefont{{G.M. D'Ariano, C. Macchiavello, P.
  Perinotti}}(2005)}]{DAriano:04}
\bibinfo{author}{\bibnamefont{{G.M. D'Ariano, C. Macchiavello, P. Perinotti}}},
  \bibinfo{journal}{Phys. Rev. A} \textbf{\bibinfo{volume}{72}},
  \bibinfo{pages}{{042327}} (\bibinfo{year}{2005}).

\bibitem[{\citenamefont{{A. Kitaev}}(1995)}]{Kitaev:95}
\bibinfo{author}{\bibnamefont{{A. Kitaev}}} (\bibinfo{year}{1995}),
  \bibinfo{note}{{quant-ph/9511026}}.

\bibitem[{\citenamefont{{R. Cleve, A. Ekert, C. Macchiavello and M.
  Mosca}}(1998)}]{Cleve:98}
\bibinfo{author}{\bibnamefont{{R. Cleve, A. Ekert, C. Macchiavello and M.
  Mosca}}}, \bibinfo{journal}{Proc. Roy. Soc. London Ser. A}
  \textbf{\bibinfo{volume}{454}}, \bibinfo{pages}{339} (\bibinfo{year}{1998}),
  \bibinfo{note}{eprint quant-ph/9708016.}

\bibitem[{\citenamefont{{D.S. Abrams and S. Lloyd}}(1999)}]{Abrams:99}
\bibinfo{author}{\bibnamefont{{D.S. Abrams and S. Lloyd}}},
  \bibinfo{journal}{Phys. Rev. Lett.} \textbf{\bibinfo{volume}{83}},
  \bibinfo{pages}{5162} (\bibinfo{year}{1999}).

\bibitem[{\citenamefont{{M.A. Nielsen and I.L. Chuang}}(2000)}]{Nielsen:book}
\bibinfo{author}{\bibnamefont{{M.A. Nielsen and I.L. Chuang}}},
  \emph{\bibinfo{title}{{Quantum Computation and Quantum Information}}}
  (\bibinfo{publisher}{{Cambridge University Press}},
  \bibinfo{address}{Cambridge, UK}, \bibinfo{year}{2000}).

\bibitem[{\citenamefont{{Joseph Emerson, Yaakov S. Weinstein, Marcos Saraceno,
  Seth Lloyd and David G. Cory}}(2003)}]{Lloyd:03}
\bibinfo{author}{\bibnamefont{{Joseph Emerson, Yaakov S. Weinstein, Marcos
  Saraceno, Seth Lloyd and David G. Cory}}}, \bibinfo{journal}{{Science}}
  \textbf{\bibinfo{volume}{{302}}} (\bibinfo{year}{2003}).

\bibitem[{\citenamefont{{M. Mohseni and D.A. Lidar}}(2006)}]{Mohseni:06}
\bibinfo{author}{\bibnamefont{{M. Mohseni and D.A. Lidar}}},
  \bibinfo{journal}{Phys. Rev. Lett.} \textbf{\bibinfo{volume}{97}},
  \bibinfo{pages}{{170501}} (\bibinfo{year}{2006}).

\bibitem[{\citenamefont{{A. Ashikhmin, A. Barg, E. Knill, S.
  Litsyn}}(1999{\natexlab{a}})}]{Ashikhmin:99a}
\bibinfo{author}{\bibnamefont{{A. Ashikhmin, A. Barg, E. Knill, S. Litsyn}}}
  (\bibinfo{year}{1999}{\natexlab{a}}), \bibinfo{note}{{quant-ph/9906126}}.

\bibitem[{\citenamefont{{A. Ashikhmin, A. Barg, E. Knill, S.
  Litsyn}}(1999{\natexlab{b}})}]{Ashikhmin:99b}
\bibinfo{author}{\bibnamefont{{A. Ashikhmin, A. Barg, E. Knill, S. Litsyn}}}
  (\bibinfo{year}{1999}{\natexlab{b}}), \bibinfo{note}{{quant-ph/9906131}}.

\bibitem[{\citenamefont{{S. L. Braunstein}}(1994)}]{Braunstein:94}
\bibinfo{author}{\bibnamefont{{S. L. Braunstein}}}, \bibinfo{journal}{Phys.
  Rev. A} \textbf{\bibinfo{volume}{49}}, \bibinfo{pages}{{49}}
  (\bibinfo{year}{1994}).

\bibitem[{\citenamefont{{V. Giovannetti, S. Lloyd and L.
  Maccone}}(2004)}]{Giovannetti:05}
\bibinfo{author}{\bibnamefont{{V. Giovannetti, S. Lloyd and L. Maccone}}},
  \bibinfo{journal}{{Science}} \textbf{\bibinfo{volume}{{306}}}
  (\bibinfo{year}{2004}).

\bibitem[{\citenamefont{{K.R. Brown, R.J. Clark and I.L.
  Chuang}}(2006)}]{Brown:06}
\bibinfo{author}{\bibnamefont{{K.R. Brown, R.J. Clark and I.L. Chuang}}},
  \bibinfo{journal}{Phys. Rev. Lett.} \textbf{\bibinfo{volume}{97}},
  \bibinfo{pages}{{050504}} (\bibinfo{year}{2006}).

\bibitem[{\citenamefont{{P. Jordan and E. Wigner}}(1928)}]{Jordan/Wigner:26}
\bibinfo{author}{\bibnamefont{{P. Jordan and E. Wigner}}},
  \bibinfo{journal}{{Z. Phys.}} \textbf{\bibinfo{volume}{{47}}}
  (\bibinfo{year}{1928}).

\bibitem[{\citenamefont{{L.-A. Wu and D. A. Lidar}}(2002)}]{Wu/Lidar:02b}
\bibinfo{author}{\bibnamefont{{L.-A. Wu and D. A. Lidar}}},
  \bibinfo{journal}{J. Math. Phys.} \textbf{\bibinfo{volume}{43}},
  \bibinfo{pages}{{4506}} (\bibinfo{year}{2002}).

\bibitem[{\citenamefont{{G. Mahler and V.A. Weberruss}}(1998)}]{Mahler:book}
\bibinfo{author}{\bibnamefont{{G. Mahler and V.A. Weberruss}}},
  \emph{\bibinfo{title}{{Quantum Networks: Dynamics of Open Nanostructures}}}
  (\bibinfo{publisher}{{Springer Verlag}}, \bibinfo{address}{{Berlin}},
  \bibinfo{year}{1998}), \bibinfo{edition}{2nd} ed.

\bibitem[{\citenamefont{{M.S. Byrd and N. Khaneja}}(2003)}]{Byrd/Khaneja:03}
\bibinfo{author}{\bibnamefont{{M.S. Byrd and N. Khaneja}}},
  \bibinfo{journal}{Phys. Rev. A} \textbf{\bibinfo{volume}{68}},
  \bibinfo{pages}{{062322}} (\bibinfo{year}{2003}), \bibinfo{note}{{ePrint
  \texttt{quant-ph/0302024}}}.

\bibitem[{\citenamefont{{Gen Kimura}}(2003)}]{Kimura}
\bibinfo{author}{\bibnamefont{{Gen Kimura}}}, \bibinfo{journal}{Phys. Lett. A}
  \textbf{\bibinfo{volume}{{314}}}, \bibinfo{pages}{{339}}
  (\bibinfo{year}{2003}).

\end{thebibliography}


\end{document}